\documentclass[fleqn,usenatbib]{mnras}

\usepackage{newtxtext,newtxmath}
\usepackage[T1]{fontenc}

\DeclareRobustCommand{\VAN}[3]{#2}
\let\VANthebibliography\thebibliography
\def\thebibliography{\DeclareRobustCommand{\VAN}[3]{##3}\VANthebibliography}

\usepackage{graphicx}	% Including figure files
\usepackage{amsmath}	% Advanced maths commands
\usepackage{xcolor}
\usepackage{units} % use units package
\usepackage{hyperref} %include url links
\title[Combining equation of state measurements]{
A scalable random forest regressor for combining neutron-star equation of state measurements: A case study with GW170817 and GW190425}

\author[Hernandez Vivanco et al.]{
Francisco Hernandez Vivanco,$^{1,2}$\thanks{E-mail: francisco.hernandezvivanco@monash.edu}
Rory Smith,$^{1,2}$
Eric Thrane$^{1,2}$
and Paul D. Lasky$^{1,2}$
\\
$^{1}$School of Physics and Astronomy, Monash University, Vic 3800, Australia\\
$^{2}$OzGrav: The ARC Centre of Excellence for Gravitational Wave Discovery, Clayton VIC 3800, Australia\\
}

\date{Accepted XXX. Received YYY; in original form ZZZ}

\pubyear{2020}

\begin{document}
\label{firstpage}
\pagerange{\pageref{firstpage}--\pageref{lastpage}}
\maketitle

\begin{abstract}
Gravitational-wave observations of binary neutron star coalescences constrain the neutron-star equation of state by enabling measurement of the tidal deformation of each neutron star. This deformation is well approximated by the tidal deformability parameter $\Lambda$, which was constrained using the first binary neutron star gravitational-wave observation, GW170817. Now, with the measurement of the second binary neutron star, GW190425, we can combine different gravitational-wave measurements to obtain tighter constraints on the neutron-star equation of state. In this paper, we combine data from GW170817 and GW190425 to place constraints on the neutron-star equation of state. To facilitate this calculation, we derive interpolated marginalized likelihoods for each event using a machine learning algorithm. These likelihoods, which we make publicly available, allow for results from multiple gravitational-wave signals to be easily combined. Using these new data products, we find that the radius of a fiducial \unit[1.4]{$M_\odot$} neutron star is constrained to \unit[$11.6^{+1.6}_{-0.9}$]{km} at 90\% confidence and the pressure at twice the nuclear saturation density is constrained to \unit[$3.1^{+3.1}_{-1.3}\times10^{34}$]{dyne/cm$^2$} at 90\% confidence. Combining GW170817 and GW190425 produces constraints indistinguishable from GW170817 alone and is consistent with findings from other works. 
\end{abstract}


%%%%%%%%%%%%%%%%%%%%%%%%%%%%%%%%%%%%%%%%%%%%%%%%%%

%%%%%%%%%%%%%%%%% BODY OF PAPER %%%%%%%%%%%%%%%%%%

\section{Introduction}

Neutron stars are some of the most compact objects found in our Universe with densities in excess of the nuclear saturation density. Such conditions cannot be simulated by Earth-based experiments and so the study of these objects offers a unique way to understand how matter behaves at supranuclear densities. The behaviour of dense matter in neutron stars is determined by the neutron star equation of state.  Gravitational-wave observations of binary neutron star coalescences allow us to constrain the neutron star equation of state by measuring the tidal deformability $\Lambda$, which is a result of the mass-quadrupole moment $Q_{ij}$ induced by the tidal field of the companion star~\citep{PhysRevD.45.1017}. The first measurement of a binary neutron star coalescence, GW170817~\citep{LSC_GW170817}, detected by LIGO and Virgo~\citep{aligo:2015,Acernese:2014}, placed the first constraints on $\Lambda$ scaled to a \unit[1.4]{$M_\odot$} neutron star to $\Lambda_{1.4} \leq 800$ at 90\% confidence, favouring compact equations of state. This observation was combined with measurements of the mass of PSR~J0348+0432~\citep{Antoniadis1233232}, to place constraints on the neutron star radius as well as the pressure inside their cores. \citet{De_2018} constrained the radius of the neutron stars in GW170817 to $\unit[8.9]{km}\leq R\leq \unit[13.2]{km}$ and \citet{Abbott_2018_GW170817_radius} constrained the radius of both neutron stars to \unit[$11.9^{+1.4}_{-1.4}$]{km} and the pressure at twice the nuclear saturation density to \unit[$3.5^{+2.7}_{-1.7}\times10^{34}$]{dyne/cm$^2$}. 

\citet{Raaijmakers_2020} combined the tidal deformabilities from GW170817 with the heaviest pulsar observed to date, PSR~J0740+6620~\citep{Cromartie:2020}, and the mass-radius measurement of pulsar PSR~J0030+0451~\citep{Riley_2019,Raaijmakers_2019,Miller_2019}. Their results are dominated by PSR~J0740+6620. 
\citet{Capano_2020} then combined GW170817 and PSR~J0030+0451, including information from low-energy nuclear theory constrained by experimental data. Their results find the tightest constraint on the neutron star equation of state, which constrain the radius of a \unit[1.4]{$M_\odot$} neutron star to \unit[$R=11.0^{+0.9}_{-0.6}$]{km} (90\% confidence). 

The second gravitational-wave measurement of a binary neutron star, GW190425~\citep{GW190425}, was detected with a signal-to-noise ratio SNR=12.9, significantly lower than GW170817. This event is interesting because the total mass of the binary is significantly heavier than any other double neutron star system~\citep{Farrow}. The fact that the binary is massive means that the tidal deformability is small and the gravitational-wave data alone cannot technically rule out that any of the objects of the binary is a black hole, though, this would be highly surprising as massive neutron stars (consistent with those of GW190425) are commonly in found in binaries with white dwarfs~\citep{Kiziltan}.\footnote{For papers seeking to explain the unusual mass of GW190425, see \cite{GW190425_origin} and \cite{Safarzadeh}.} Despite the low SNR of GW190425, it was possible to map the tidal deformabilities of GW170817 to the mass scale of GW190425 in order to constrain the equation of state \citep{GW190425}, but the results are dominated by the prior, meaning that the data are not informative enough to place tighter constraints on the equation of state.

Neutron star-black hole coalescences can also potentially constrain the neutron star equation of state. A candidate for such an event is GW190814~\citep{GW190814} which is the result of a merger of a \unit[$23.2^{+1.1}_{-1.0}$]{$M_\odot$} black hole with a \unit[$2.59^{+0.08}_{-0.09}$]{$M_\odot$} compact object. It is not clear whether the compact object is the heaviest neutron star or the lightest black hole observed to date. The tidal deformability of the low mass object is uninformative and no electromagnetic counterpart was observed,  which is consistent with a black hole or a neutron star due to the extreme mass ratio and distance of this event~\citep{fernndez_2020,morgan_2020}. However, we can use the maximum neutron star mass ($m_\text{TOV}$) to determine the nature of this object. If the mass of the compact object is greater than $m_\text{TOV}$, we can assume it is a black black hole. Current constraints on the neutron star maximum mass from GW170817 tidal-deformability measurements imply $m_{\rm TOV}\lesssim\unit[2.3]{M_\odot}$~\citep{Lim_2019,Essick:2020}, supporting the conclusion that the $m=\unit[2.59^{+0.08}_{-0.09}]{M_\odot}$ secondary of GW190814 is too massive to be a neutron star.  This claim is further strengthened when the GW170817 constraint is combined with equation of state inference results from terrestrial heavy-ion experiments~\citep{Fattoyev_2020}.

With the increasing number of binary neutron star measurements from gravitational-wave observations and electromagnetic observations, it is important moving forward to have a framework that allows the community to easily combine different measurements constraining the neutron-star equation of state. 
In \cite{Hernandez_Vivanco_2019}, we highlighted the technical challenges associated with equation-of-state inference using multiple gravitational-wave events.
We pointed out that the usual method of releasing posterior samples is not conducive to equation-of-state inference because inference calculations require the computation of line integrals, which in general do not pass through any of the posterior samples.
We proposed a new paradigm, which makes use of machine-learning representations of marginal likelihood surfaces.
Similar to our method, the work presented in~\citet{Wysocki_2020} solves the problem of combining gravitational-wave observations to constrain the equation of state by interpolating the marginalised likelihood using either random forest or Gaussian process interpolation. 
Their method is used to infer the merger rate and mass distribution of neutron stars in addition to the neutron-star equation of state.
See also \cite{Lackey_2015,Agathos_2015} for other approaches to stacking gravitational-wave signals for equation-of-state inference.
For a different approach to calculating marginal likelihoods, see \cite{Pankow_2015,Lange_2018}, which use adaptive mesh refinement to calculate marginal likelihoods on a mesh grid as in~\citet{Abbott_2018_GW170817_radius}.

In this paper, we build on  \citet{Hernandez_Vivanco_2019} to present constraints on the neutron star equation of state obtained from combining the first two binary neutron star gravitational-wave observations, GW170817 and GW190425. We do not include GW190814 in our analysis because it is unlikely that the compact object is a neutron star and, if it is a neutron star, the tidal deformability is uninformative and does not provide any additional constraints to the neutron star equation of state~\citep{GW190814}.  While combining data from GW170817 and GW190425, we calculate marginalised likelihoods of GW170817 and GW190425 using a machine learning algorithm consisting of a random forest regressor. We make these data products publicly available. This form of data release is useful for equation of state measurements from multiple measurements. 

The advantage of the marginalised likelihoods calculated in this study is that they are continuous and can be evaluated at any point of the $(m,\Lambda)$ plane supported by the posterior distributions of GW170817 and GW190425. (This is helpful for evaluating the aforementioned line integrals required for equation-of-state inference.) Additionally, we can adaptively refine the interpolation by calculating the interpolated likelihood with greater density in the intrinsic parameters depending on the data, which allows us to achieve the necessary interpolation accuracy for whatever calculation may be required.

This paper is organised as follows. In Sec.~\ref{sec:method}, we give an overview of the method we use to combine gravitational-wave observations and explain why interpolating the likelihood distribution solves the problem of combining events using hierarchical Bayesian inference. In Sec.~\ref{sec:data_release} we explain how to use the interpolated likelihoods released in this study. In Sec.~\ref{sec:results} we present constraints on the equation of state using the interpolated likelihoods. In Sec.~\ref{sec:discussion} we discuss our results and we conclude in Sec.~\ref{sec:conclusions}.

\section{Method}\label{sec:method}
We follow the method we introduced in~\citet{Hernandez_Vivanco_2019}, which presents a solution to the ``stacking problem'' found in hierarchical Bayesian inference. We briefly explain how our method works in practise as follows. 

We start by writing Bayes theorem, where our aim is to obtain a posterior distribution $p(\Upsilon | \vec{d})$ on the hyper-parameters $\Upsilon$ that define the neutron star equation of state.
\begin{align}\label{eq:bayes_theorem}
    p(\Upsilon | \vec{d})
    = \frac{{\cal L}(\vec{d}|\Upsilon)\pi(\Upsilon)}{\mathcal{Z}_\Upsilon} .
\end{align}
The posterior $p(\Upsilon | \vec{d})$ depends on the hyper-likelihood $\mathcal{L}(\vec{d}|\Upsilon)$, the hyper-prior $\pi(\Upsilon)$ and the evidence  $\mathcal{Z}_\Upsilon$. Here, the likelihood $\mathcal{L}_\text{tot}$ is defined by

\begin{align} 
    \label{eq:likelihood_product}
    {\cal L}_\text{tot}(\vec{d}|\Upsilon)
    = \prod_i^N \int d\theta_i
    {\cal L}(d_i|\theta_i)
    \pi(\theta_i | \Upsilon),
\end{align}
where $N$ are the number of gravitational-wave events that we combine and $\theta$ are the parameters that model the properties of a binary neutron star coalescence. Our method can be extended to account for X-ray observations, e.g. by NICER \citep{Miller_2019}, by adding another term in the likelihood defined in Equation~(\ref{eq:likelihood_product}), which would place additional constraints on the mass and radius of neutron stars. However, in this study we focus only on gravitational-wave observations.

It can be shown that the multi-detector likelihood distribution $\mathcal{L}_\text{tot}(\vec{d}|\Upsilon)$  can be expressed as ~\citep[e.g.][]{Thrane_2019}

\begin{align}\label{eq:recycling}
    {\cal L}_\text{tot}(\vec{d}|\Upsilon) = &
    \prod_i^N \frac{\mathcal{Z}_{\emptyset}^i}{n_i} 
    \sum_k^{n_i} \frac{\pi(\theta_i^k|\Upsilon)}{\pi(\theta_i^k|\emptyset)},
\end{align}
where $n_i$ are the posterior samples obtained from running parameter estimation on individual events using an initial prior $\pi(\theta_i^k|\emptyset)$ and $\pi(\theta_i^k|\Upsilon)$ is the hyper-prior that depends on hyper-parameters $\Upsilon$ that model the neutron star equation of state.

The stacking problem occurs when we try to combine posterior samples to probe deterministic curves represented in the hyper-prior $\pi(\theta_i^k|\Upsilon)$, i.e, curves that are infinitely thin instead of having a probability distribution spanning over an area of the parameter space. Since the equation of state is defined by a curve in the $(\Lambda,m)$ plane, we find that no posterior sample will fall exactly on the curve defined by the equation of state and Equation (\ref{eq:recycling}) evaluates to zero. 

We solve this issue in~\citet{Hernandez_Vivanco_2019} by interpolating the marginalised likelihood for each gravitational-wave observation. This is different to using kernel density estimation (KDE) to represent posterior samples, as in ~\citep[e.g.][]{Lackey_2015,Raaijmakers_2020}, because KDEs perform density estimation whereas likelihood interpolation is a direct surrogate for the underlying function.
The marginalised likelihood depends on the intrinsic parameters $\omega=(m_1,m_2,\Lambda_1,\Lambda_2)$ that determine the neutron star equation of state. By marginalising the likelihood, we can rewrite the total likelihood defined in Equation (\ref{eq:likelihood_product}) as
\begin{align}\label{eq:total_interpolated_likelihood}
    {\cal L}_\text{tot}(\vec{d}|\Upsilon) = & \prod_i^N \int d\omega_i
    {\cal L}_\kappa^\text{int}(d_i|\omega_i)
    \pi(\omega_i|\Upsilon) ,
\end{align}
where ${\cal L}_\kappa^\text{int}(d_i|\omega_i)$ is the interpolated likelihood marginalised over the parameters $\kappa$ that are not in $\omega$. 

The interpolated likelihood is obtained by running parameter estimation with parameters $\omega$ fixed at random interpolation points $\omega_i$, where we obtain evidences $\mathcal{Z}_i$ that effectively represent  the marginalised likelihood evaluated at $\omega_i$. The data generated during this step is used to train a random forest regressor~\citep{Breiman:2001} to predict the marginalised likelihood at any point $\mathcal{L}(d|\omega_i)$. We refer to this step as ``second-stage parameter estimation''. By working with the interpolated likelihood defined in Equation (\ref{eq:total_interpolated_likelihood}), we do not work with posterior samples at any point and we avoid the issue found in Equation (\ref{eq:recycling}).

\subsection{Second-stage parameter estimation}
To obtain the interpolation likelihood distributions defined in Equation (\ref{eq:total_interpolated_likelihood}), we run parameter estimation by fixing random intrinsic parameters $\omega_i=(\mathcal{M}_i,q_i,\Lambda_{1,i},\Lambda_{2,i})$, where $\mathcal{M}$ is the chirp mass and $q$ is the mass ratio, to evaluate the marginalised likelihood distribution evaluated at $\omega_i$. We refer to this step as ``second-stage parameter estimation''. The values of  $\omega_i$ are chosen from the posterior distributions of each event as well as random points from the prior. We run second-stage parameter estimation with \textsc{Bilby}~\citep{Ashton_2019} using the \textsc{Dynesty} sampler~\citep{Dynesty}. There are some subtleties when running second-stage parameter estimation which we detail below.

The duration of binary neutron star signals is in the order of minutes. Running parameter estimation of binary neutron star inspirals is therefore more computationally expensive than lower-duration events such as binary black hole coalescences. One of the solutions to this problem is to use reduced-order models (ROM)~\citep{Smith_2016}. The key idea of this method is to remove redundant evaluations of the waveform at some frequency bins, which enables the evaluation of significantly cheaper Bayesian probability distributions using reduced order quadrature (ROQ) integration. This can accelerate Bayesian parameter estimation by as much as a factor of 300  compared to running parameter estimation using the full waveform approximant.

In our analysis,  we run second-stage parameter estimation on GW190425 using an ROQ implementation of the precessing-spin waveform approximant \texttt{IMRPhenomPv2\_NRTidal}~\citep{Khan:2016,Baylor_2019} starting at a frequency \unit[$f_\text{min} = 19.4$]{Hz}. Similarly, we analyse GW170817 using an ROQ implementation of the spin-aligned waveform approximant \texttt{IMRPhenomD\_NRTidal}~\citep{Husa:2016,Dietrich:2017}  starting at a frequency \unit[$f_\text{min} = 32$]{Hz}. We do not use the same waveform approximant for GW170817 and GW190425 because we do not currently have an ROQ implementation of the \texttt{IMRPhenomPv2\_NRTidal} approximant spanning over the chirp mass values defined by the GW170817 prior. In both cases, we assume a low-spin prior as detailed in Table~\ref{tab:prior_gw170817} and Table~\ref{tab:prior_gw190425}. Note that the minimum frequency at which we start the analyses of both events is different. The reason why we analyse GW170817 from \unit[32]{Hz} is related to discontinuities in the waveform that break the requirement for the greedy basis finding algorithm defined in~\citet{Smith_2016}, that require the model be smooth. However, when we analyse GW170817 from \unit[32]{Hz}, contrary to \unit[$23$]{Hz} as in~\citet{Romero_Shaw_2020}, we lose a signal-to-noise ratio of $\sim 1$. This does not affect the information about the tidal deformabilities, consistent with~\citet{Harry_2018}, but the chirp mass posterior distribution changes from $\mathcal{M}_{\text{23 Hz}} = 1.19755^{+0.00012}_{-0.00011}$ to $\mathcal{M}_{\text{32 Hz}} = 1.19751^{+0.00020}_{-0.00017}$ (90\% confidence) and the lower mass ratio limit changes from $q_\text{23 Hz}=0.759$ to $q_\text{32 Hz}=0.750$ (90\% confidence).

\begin{table}
    \begin{tabular}{ccccc}
    Parameter             &Unit         & Prior     & Minimum & Maximum \\
    \hline
    $\mathcal{M}$         &$M_\odot$    & Uniform   & 1.18    & 1.21   \\
    $q$                   &-            & Uniform   & 0.125   & 1       \\
    $\Lambda_1$, $\Lambda_2$  &-        & Uniform   & 0       & 5000  \\
    $a_1$, $a_2$         & -           & Uniform   & 0   & 0.05    \\
    %RA                    &rad.         & Uniform   & 0       & $2\pi$  \\
    %DEC                   &rad.         & Cos       & $-\pi/2$& $\pi/2$ \\
    $\cos(\theta_{jn})$   &-            & Uniform   & -1      & 1       \\
    $\psi$                &rad.         & Uniform   & 0       & $\pi$   \\
    $\phi$                &rad.         & Uniform   & 0       & $2\pi$  \\
    $d_L$                 &Mpc          & Comoving & 1       & 75     \\
    \end{tabular}
    \caption{Prior distributions used in the analysis of GW170817. In this table, $\mathcal{M}$ is the chirp mass, $q$ is the mass ratio, $\Lambda_{1,2}$ are the tidal deformabilities, $a_{1,2}$ are the spin magnitudes, $\theta_{jn}$ is the inclination angle, $\psi$ is the polarization angle, $\phi$ is the binary phase and $d_L$ is the luminosity distance. We fix the right ascension (RA) and declination (DEC) to 3.44616 and -0.408084 degrees respectively consistent with electromagnetic observations and we use an aligned-spin prior. } 
    \label{tab:prior_gw170817}
\end{table}

\begin{table}
    \begin{tabular}{ccccc}
    Parameter             &Unit         & Prior     & Minimum & Maximum \\
    \hline
    $\mathcal{M}$         &$M_\odot$    & Uniform   & 1.485   & 1.49    \\
    $q$                   &-            & Uniform   & 0.125   & 1       \\
    $\Lambda_1$, $\Lambda_2$  &-        & Uniform   & 0       & 5000    \\
    $a_1$, $a_2$          & -           & Uniform   & 0       & 0.05    \\
    $\theta_1$, $\theta_2$ & rad          & Sin       & 0       & $\pi$   \\
    $\phi_{12}$, $\phi_{jl}$ & rad        & Uniform   & 0       & $2\pi$  \\
    RA                    &rad.         & Uniform   & 0       & $2\pi$  \\
    DEC                   &rad.         & Cos       & $-\pi/2$& $\pi/2$ \\
    $\cos(\theta_{jn})$   &-            & Uniform   & -1      & 1       \\
    $\psi$                &rad.         & Uniform   & 0       & $\pi$   \\
    $\phi$                &rad.         & Uniform   & 0       & $2\pi$  \\
    $d_L$                 &Mpc          & Comoving  & 1      & 500     \\
    \end{tabular}
    \caption{Prior distributions used in the analysis of GW190425. The parameters are the same as Table~\ref{tab:prior_gw170817} with the difference that we use a precessing-spin prior. } 
    \label{tab:prior_gw190425}
\end{table}

\section{Marginalised likelihood fits}\label{sec:data_release}
\subsection{Validation}
The key idea of our work is to obtain marginalised likelihood distributions for individual gravitational-wave observations to avoid the stacking problem. We interpolate the likelihood distribution with a random forest regressor~\citep{Breiman:2001} using the Python package \textsc{Scikit-learn}~\citep{scikit-learn}. A random forest is a bagging algorithm that combines the results from random decision trees to make a prediction. Each tree is trained individually and there is no interaction between each decision tree during training. The results are obtained by averaging the outcomes of each tree which reduces the risk of over-fitting~\citep{Biau_2016}. In this study, we use 50 decision trees to train our model. Once the model is trained, we use it to predict the marginalised likelihood $\mathcal{L}$, given intrinsic parameters $w=(\mathcal{M},q,\Lambda_1,\Lambda_2)$. We generate $\sim 6\times10^3$ interpolation points for GW190425 and $\sim 11\times10^3$ interpolation points for GW170817. We use 90\% of this data for training and 10\% for testing.

\begin{figure}
	\includegraphics[width=\columnwidth]{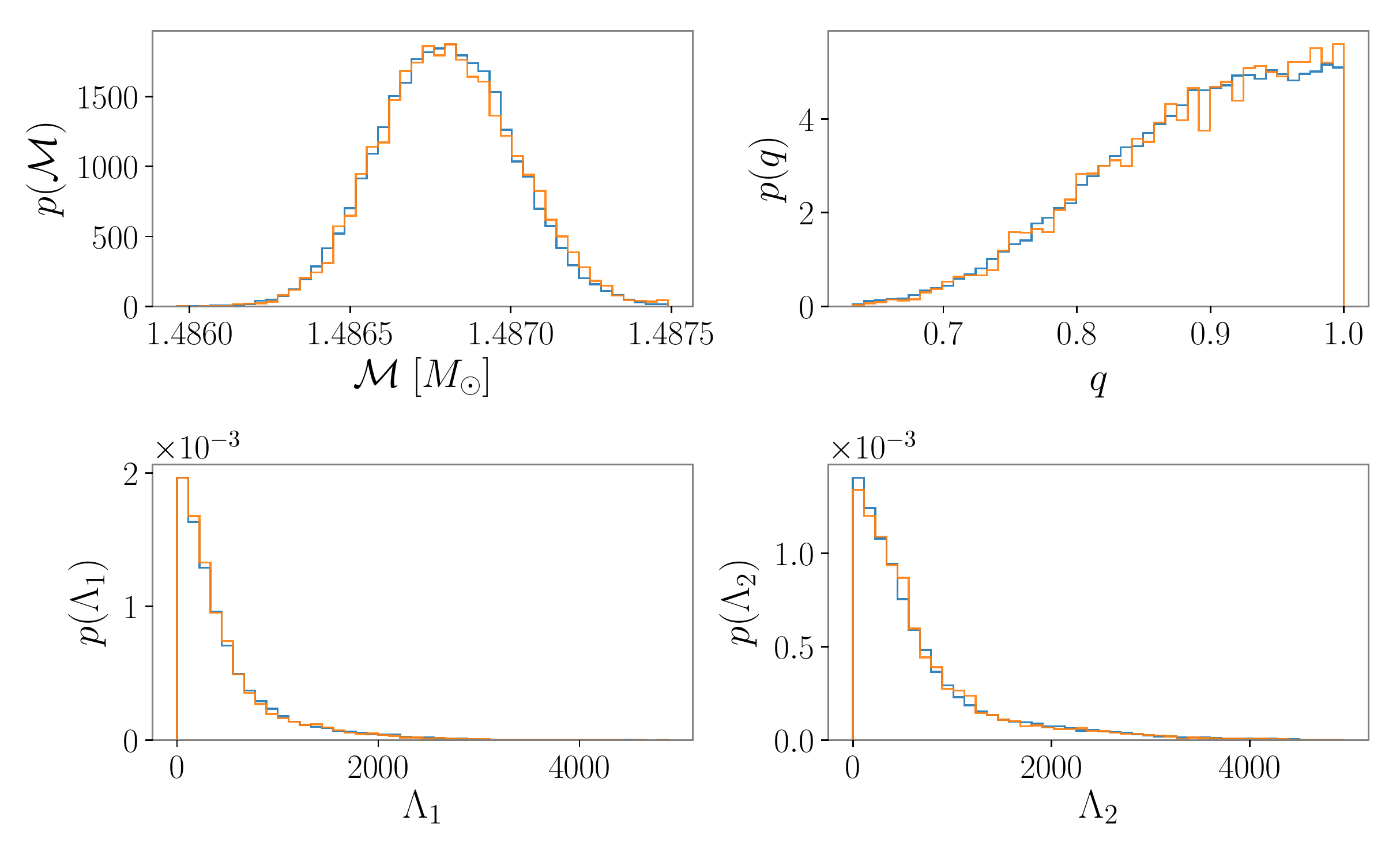}
    \caption{Posterior distributions comparing the original posterior samples from GW190425, shown in blue, with the samples obtained from the interpolated likelihood, shown in orange.}
    \label{fig:GW190425}
\end{figure}

\begin{figure}
	\includegraphics[width=\columnwidth]{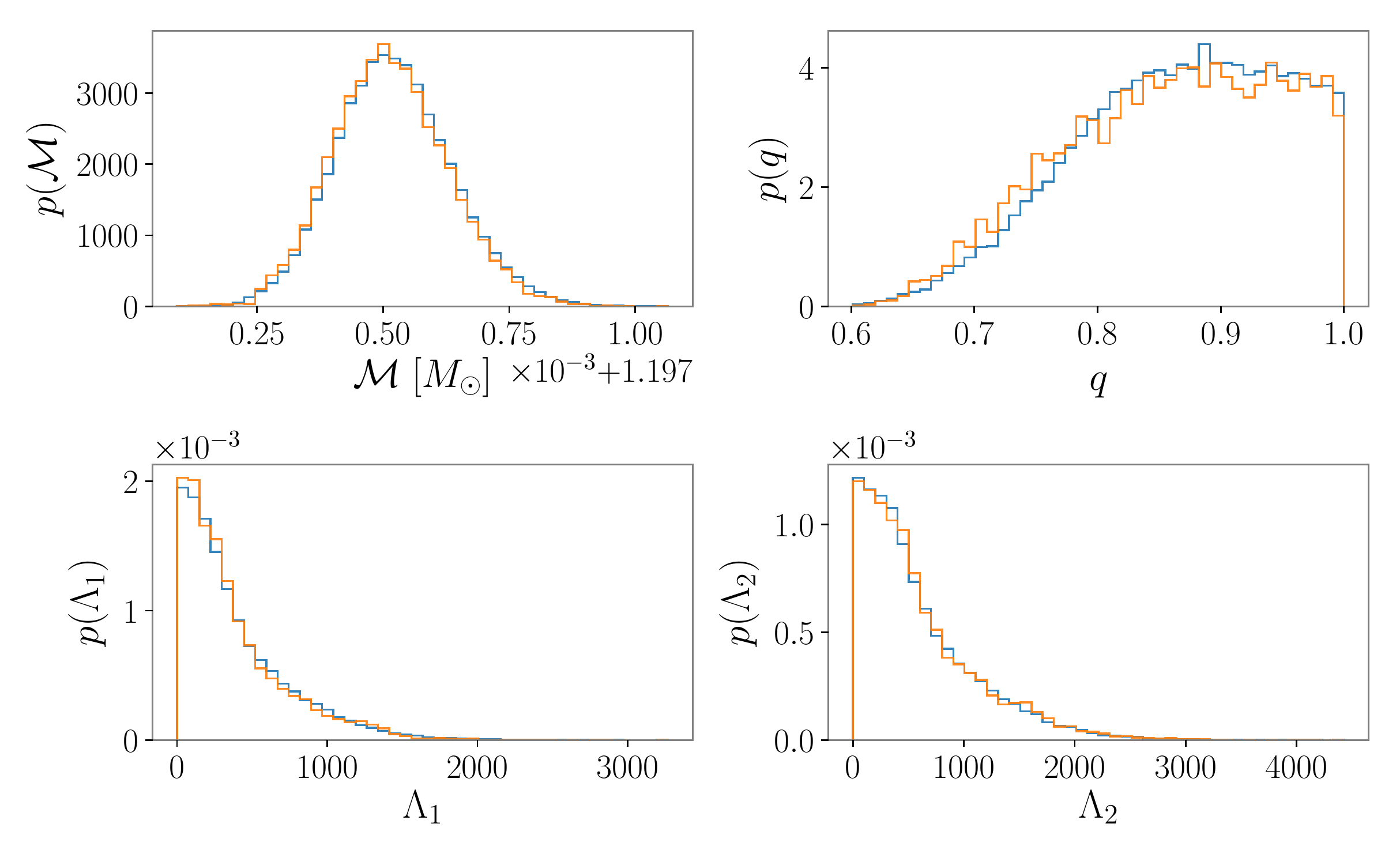}
    \caption{Posterior distributions comparing the original posterior samples from GW170817, shown in blue, with the samples obtained from the interpolated likelihood, shown in orange.}
    \label{fig:GW170817}
\end{figure}

To check if an interpolated marginalised likelihood reproduces the original posterior, we sample the interpolated marginalised likelihood and check if the posterior distributions are consistent. We do this for GW170817 and GW190425. The results are shown in Fig.~\ref{fig:GW190425} and Fig.~\ref{fig:GW170817}, where we see that the interpolated likelihoods accurately reproduce the original posteriors. To quantitatively determine if both distributions are similar, we use the Jensen–Shannon (JS) divergence~\citep{Lin_1991}.  In~\citet{Romero_Shaw_2020}, it was found that posteriors with JS divergence values \unit[$\gtrsim 0.002$]{bit} are statistically significant. This value is somewhat arbitrary: \citet{Romero_Shaw_2020} compared a large set of posteriors from multiple gravitational-wave events and demonstrated that a JS divergence of $\lesssim 0.002$ bit results in consistent posteriors, i.e., the median and confidence intervals are essentially the same. For consistency, in this paper, we adopt the threshold value of 0.002 bit proposed in \citet{Romero_Shaw_2020} to compare the original posteriors and the posteriors obtained from sampling the interpolated likelihood. We calculate the JS values for all four intrinsic parameters $\omega=(\mathcal{M},q,\Lambda_1,\Lambda_2)$ and find that the maximum JS values for GW190425 and GW170817 are \unit[0.001]{bit} and \unit[0.002]{bit} respectively, validating the accuracy of our interpolated likelihoods. 

\subsection{Data release}
We make the GW170817 and GW190425 interpolated marginalised likelihoods publicly available. These likelihoods can be used to reproduce the posteriors shown in Fig.~\ref{fig:GW190425} and Fig.~\ref{fig:GW170817}. This form of data release is potentially more useful than releasing posterior samples alone, as is usually done~\citep[e.g.][]{GWTC_1_LIGO,De_2019,gwosc,Romero_Shaw_2020}. While posterior samples can be used in hierarchical Bayesian inference as long as Equation~\ref{eq:recycling} can be evaluated ~\citep[e.g.][]{Talbot_2018,Smith_2020}, we cannot use posterior samples alone to constrain the neutron star equation of state without relying on KDE-based methods, as explained in Sec.~\ref{sec:method}. An interpolated likelihood, on the other hand, can be evaluated at any point of the $(m,\Lambda)$ parameter space and is therefore ideal for use when sampling equation of state hyper-parameters. Moreover, our trained models are fast to evaluate, with a single likelihood evaluation taking in the order of \unit[$\sim 6$]{ms}.

The interpolated marginalised likelihoods can be found in our neuTrOn stAr STacking package, \textsc{Toast}\footnote{The source code, interpolation points and examples are  available in \url{https://git.ligo.org/francisco.hernandez/toast}}. Our Python package uses a random forest regressor to predict the log likelihood. However, other interpolation methods could improve the accuracy of a random forest regressor.
Therefore the interpolation points of GW170817 and GW190425 are also publicly available.

\section{Case study: combined equation of state measurement on GW170817 and GW190425}\label{sec:results}
We carry out hierarchical Bayesian inference following the method described in ~\citet{Hernandez_Vivanco_2019}.
We combine data from GW170817 and GW190425 assuming that both events are the result of binary neutron star coalescences following the same equation of state. We  assume the piecewise polytrope parametrisation of the equation of state~\citep{Read_2009}, which models pressure $p$ as a function of density $\rho$ with three different polytropes. Each polytrope has the form
\begin{align}
    p = K\rho^{\Gamma}.
\end{align}
To fully determine the equation of state with three polytropes, we use four hyper-parameters $\Upsilon=\{\log_{10} p_0,\Gamma_1,\Gamma_2,\Gamma_3\}$, where  $\log_{10} p_0$ is a reference pressure and $\Gamma_i$ represents the slope of each polytrope. To convert the gravitational-wave measurable parameters $(m,\Lambda)$ to $(p,\rho)$, we solve the Tolman-Volkoff-Oppenheimer (TOV) equations along with the second Love number $k_2$~\citep{Lattimer_2001,Hinderer_2008} using the LIGO Algorithm Library LALSuite~\citep{lalsuite}.

While sampling the piecewise polytrope hyper-parameters $\Upsilon$, we impose three conditions:
\begin{enumerate}
    \item The equation of state must be monotonic, i.e. $dp/d\rho \geq 0$.
    \item We require all samples to satisfy that \unit[$m_\text{TOV} \geq 1.97$]{$M_\odot$}, consistent with pulsars PSR~J0348+0432~\citep{Antoniadis1233232} and PSR~J0740+6620~\citep{Cromartie:2020}.
    \item The speed of sound $v_s$ should not exceed the speed of light $c$. In practice we, set the restriction to $v_s \leq 1.1c$ due to errors introduced by the equation of state parametrisation, as in ~\citet{Lackey_2015,Carney_2018}.
\end{enumerate}

\begin{figure}
	\includegraphics[width=\columnwidth]{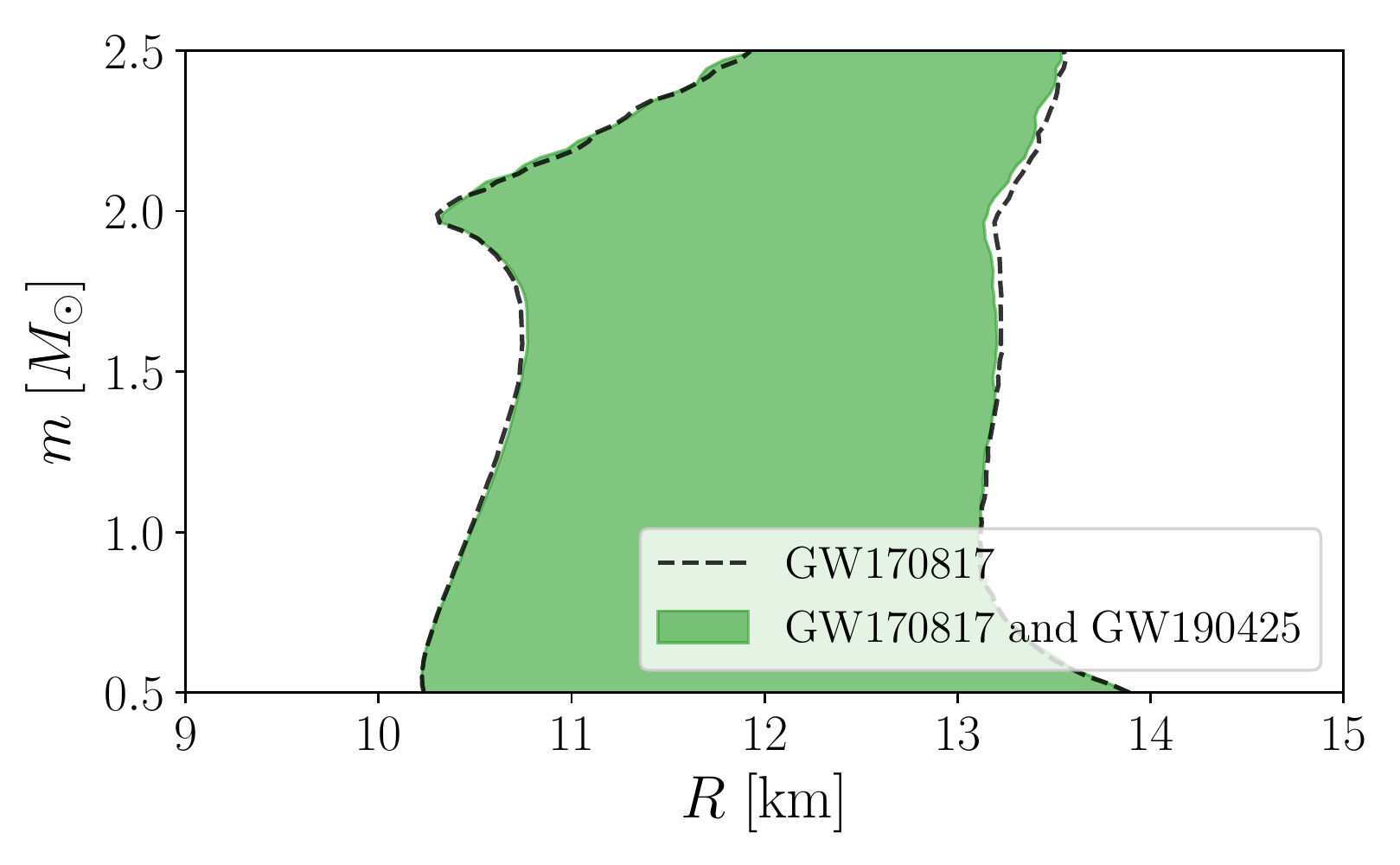}
	\includegraphics[width=\columnwidth]{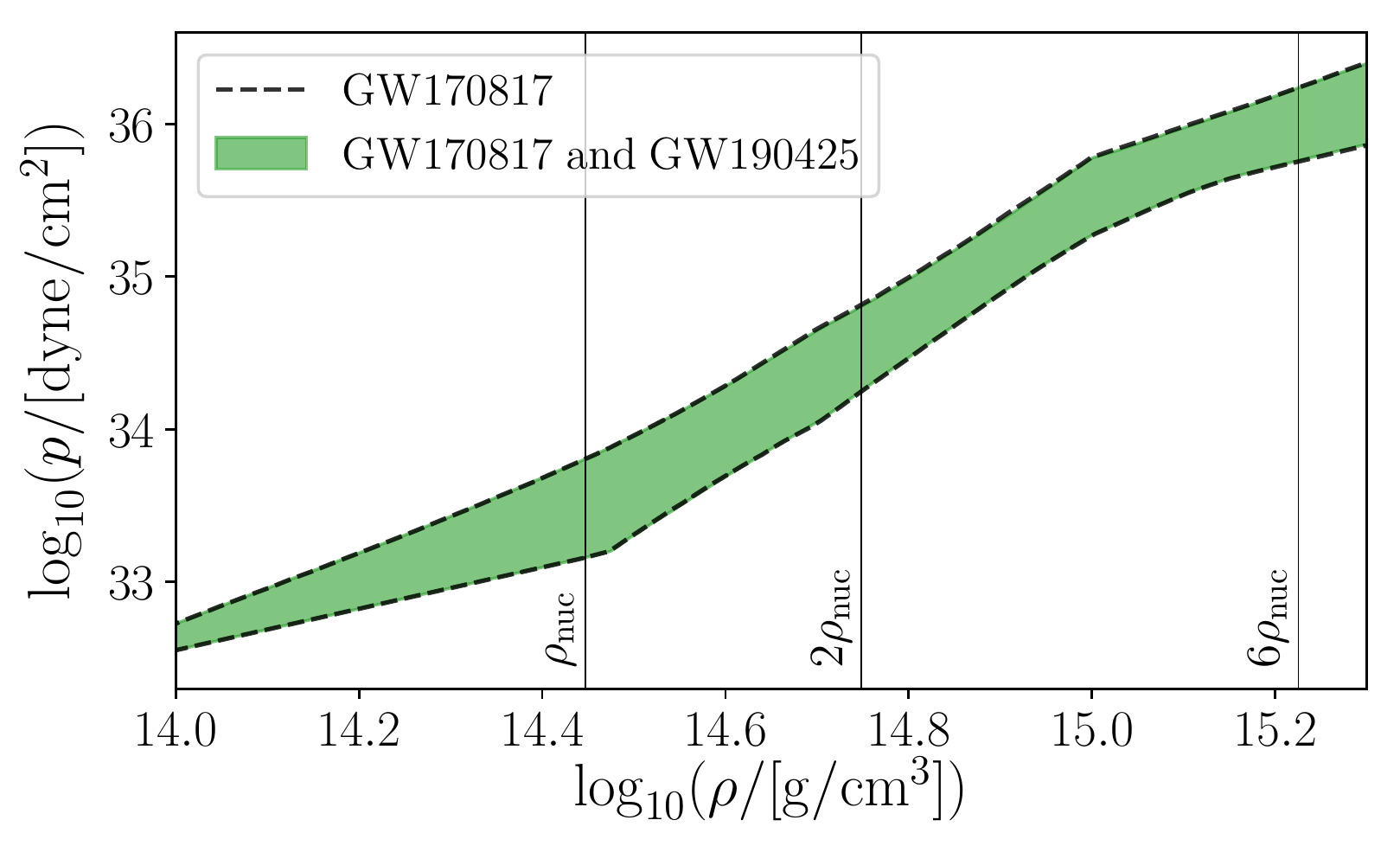}
    \caption{ 90\% confidence intervals of the posterior distributions  of $m(R)$ and $p(\rho)$ obtained by combining GW170817 and GW190425. The neutron star radius at a mass of \unit[1.4]{$M_\odot$} is constrained to \unit[$11.6^{+1.6}_{-0.9}$]{km} at 90\% confidence and the pressure at $2\rho_\text{nuc}$ ($6\rho_\text{nuc}$) is constrained to \unit[$3.1^{+3.1}_{-1.3}\times10^{34}$]{dyne/cm$^2$} (\unit[$8.3^{+8.6}_{-2.6}\times10^{35}$]{dyne/cm$^2$}) at 90\% confidence. Our results are dominated by GW170817. }
    \label{fig:posteriors}
\end{figure}

Using the conditions detailed above, we combine GW170817 and GW190425 by sampling the hyper-parameters $\Upsilon$ using the Bayesian inference library for gravitational-wave astronomy \textsc{Bilby}~\citep{Ashton_2019}. We use the posterior samples of $\Upsilon$ to obtain posterior samples for $p(\rho)$ and $m(R)$. Our results are shown in Fig.~\ref{fig:posteriors}. We find that the neutron star radius at a mass of \unit[1.4]{$M_\odot$} is constrained to \unit[$11.6^{+1.6}_{-0.9}$]{km} at 90\% confidence and the pressure at $2\rho_\text{nuc}$ ($6\rho_\text{nuc}$) saturation density is constrained to \unit[$3.1^{+3.1}_{-1.3}\times10^{34}$]{dyne/cm$^2$} (\unit[$8.3^{+8.6}_{-2.6}\times10^{35}$]{dyne/cm$^2$}) at 90\% confidence. The results obtained by combining GW170817 and GW19045 are indistinguishable from the constraints of GW170817 alone at the precision given and are consistent with ~\citet{Abbott_2018_GW170817_radius,GW190425}.

\section{Discussion}\label{sec:discussion}
\citet{Abbott_2018_GW170817_radius} showed that GW170817 constrains the pressure at $2\rho_\text{nuc}$ ($6\rho_\text{nuc}$) to \unit[$3.5^{+2.7}_{-1.7}\times 10^{34}$]{dyne/cm$^2$} (\unit[$9.0^{+7.9}_{-2.6}\times 10^{35}$]{dyne/cm$^2$}) at 90\% confidence.
Similarly, GW190425 constrains the pressure at $2\rho_\text{nuc}$ ($6\rho_\text{nuc}$) to \unit[$5.9^{+4.9}_{-3.1}\times 10^{35}$]{dyne/cm$^2$} (\unit[$9.4^{+13.3}_{-4.2}$]{dyne/cm$^2$})  at 90\% confidence~\citep{GW190425}. Both of these results are obtained assuming the binaries consist of two neutron stars with the same equation of state, assuming a low-spin prior and setting a maximum neutron-star mass prior to \unit[$m_\text{TOV} \geq 1.97$]{$M_\odot$}. 

Using our method and the same assumptions for the analysis of GW170817 and GW190425, we find that the pressure at $2\rho_\text{nuc}$ ($6\rho_\text{nuc}$) is constrained to \unit[$3.1^{+3.1}_{-1.3}\times10^{34}$]{dyne/cm$^2$} (\unit[$8.3^{+8.6}_{-2.6}\times10^{35}$]{dyne/cm$^2$}) at 90\% confidence, consistent with the results presented in \citet{LSC_GW170817,GW190425}. Furthermore, we place limits on the radius of a \unit[1.4]{$M_\odot$} neutron star to \unit[$11.6^{+1.6}_{-0.9}$]{km} at 90\% confidence.  \citet{De_2018} infer a neutron star radius in the range $8.9 \text{ km} \leq R \leq 13.2 \text{ km}$ from GW170817, assuming different mass priors and a piecewise polytrope parametrization of the equation of state. Similarly, \citet{Essick:2020} find that the radius at a mass of \unit[$1.4$]{$M_\odot$} is constrained to \unit[$10.86^{+2.04}_{-1.86}$]{km} (\unit[$12.51^{+1.00}_{-0.88}$]{km}) with nonparametric priors loosely (tightly) constrained from equations of state found in the literature~\citep{Landry_2019}. Finally, \citet{Dietrich_2020} constrain the radius of a \unit[1.4]{$M_\odot$} neutron star to \unit[$11.74^{+0.98}_{-0.79}$]{km} (90\% confindence) by combining GW170817 with its electromagnetic counterparts GRB170817A and AT2017gfo along with NICER measurements, GW190425 and nuclear-physics constraints. Our results are consistent with~\citet{Abbott_2018_GW170817_radius,De_2018,GW190425,Essick:2020,Dietrich_2020}.

Although the results from ~\citet{Abbott_2018_GW170817_radius,De_2018,Essick:2020} are obtained by analysing GW170817 alone, these are consistent with our results obtained from combining GW170817 and GW190425 because the posteriors are dominated by GW170817. This is consistent with ~\citet{Hernandez_Vivanco_2019}, where it was found that the constraints on the equation of state are dominated by events with SNR~$\gtrsim 20$. 

\section{Conclusions}~\label{sec:conclusions}
We constrain the neutron star equation of state by combining the gravitational-wave measurements of GW170817 and GW190425. To do so, we calculate interpolated marginalised likelihoods for both events using a random forest regressor. The interpolated likelihoods of GW170817 and GW190425 are made public and we argue that this form of data release is more useful than releasing posterior samples alone. 
Using the  interpolated likelihoods calculated in this study, we find that the radius of a \unit[1.4]{$M_\odot$} neutron star is constrained to \unit[$11.6^{+1.6}_{-0.9}$]{km} at 90\% confidence and the pressure at $2\rho_\text{nuc}$ ($6\rho_\text{nuc}$) is constrained to \unit[$3.1^{+3.1}_{-1.3}\times10^{34}$]{dyne/cm$^2$} (\unit[$8.3^{+8.6}_{-2.6}\times10^{35}$]{dyne/cm$^2$}) at 90\% confidence, consistent with results found in the literature.

\section*{Acknowledgements}

This work is supported through Australian Research Council Grant No. CE170100004, No. FT150100281, No. FT160100112, and No. DP180103155.
F.H.V. is supported through the Monash Graduate Scholarship (MGS).
The results presented in this manuscript were calculated using the computer clusters at California Institute of Technology and Swinburne University of Technology (OzSTAR). The authors are grateful for computational resources provided by the LIGO Laboratory and supported by National Science Foundation Grants No. PHY-0757058 and No. PHY-0823459.
This is LIGO Document No.P2000299.

%%%%%%%%%%%%%%%%%%%%%%%%%%%%%%%%%%%%%%%%%%%%%%%%%%



\bibliographystyle{mnras}
\bibliography{bibliography} % if your bibtex file is called example.bib

%\begin{thebibliography}{99}
%\end{thebibliography}

%%%%%%%%%%%%%%%%%%%%%%%%%%%%%%%%%%%%%%%%%%%%%%%%%%

%%%%%%%%%%%%%%%%% APPENDICES %%%%%%%%%%%%%%%%%%%%%

%\appendix



%%%%%%%%%%%%%%%%%%%%%%%%%%%%%%%%%%%%%%%%%%%%%%%%%%

% Don't change these lines
\bsp	% typesetting comment
\label{lastpage}
\end{document}